# Inducing Superconducting Correlation in Quantum Hall Edge States

Gil-Ho Lee[1], Ko-Fan Huang[1], Dmitri K. Efetov[2], Di S. Wei[1], Sean Hart[1], Takashi Taniguchi[3], Kenji Watanabe[3], Amir Yacoby[1], and Philip Kim[1,*]

[1]*Department of Physics, Harvard University, Cambridge, Massachusetts 02138, USA*
[2]*Department of Electrical Engineering, M.I.T., Cambridge, MA 02138*
[3]*National Institute for Materials Science, Namiki 1-1, Tsukuba, Ibaraki 305-0044, Japan*
*Correspondence and requests for materials should be addressed to P.K. (email: pkim@physics.harvard.edu).

The quantum Hall (QH) effect supports a set of chiral edge states at the boundary of a 2-dimensional electron gas (2DEG) system. A superconductor (SC) contacting these states can induce correlations of the quasi-particles in the dissipationless 1D chiral QH edge states. If the superconducting electrode is narrower than the superconducting coherence length, the incoming electron is correlated to the outgoing hole along the chiral edge state by the Andreev process[1-3] across the SC electrode. In order to realise this crossed Andreev conversion (CAC)[4-7], it is necessary to fabricate highly transparent and nanometer-scale superconducting junctions to the QH system. Here we report the observation of CAC in a graphene QH system contacted with a nanostructured NbN superconducting electrode. The chemical potential of the edge states across the SC electrode exhibits a sign reversal, providing direct evidence of CAC. This hybrid SC/QH system is a novel route to create isolated non-Abelian anyonic zero modes, in resonance with the chiral QH edge[7-12].

Inducing superconducting correlations via the proximity effect into a 2DEG in the QH regime has been a long standing challenge and has attracted renewed attentions due to its promise for realising non-Abealian zero modes[13-15]. Unlike conventional conductors, the 2DEG can exhibit an insulating incompressible bulk electronic state under perpendicular quantizing magnetic fields. In this QH regime, the conduction of electric charge occurs only along the edges via 1D chiral edge states, which the SC can make contacts to. In order to realise the hybrid system of QH and SC, the upper critical field of the SC needs to be high enough such that Cooper pairs in the SC are correlated mostly to the quasi-particles in well-developed 1D QH edge states. The experimental realisation of such hybrid systems often encounters challenges in semiconductor 2DEGs due to the formation of large Schottky barriers at the SC/semiconductor interfaces[16]. Graphene is a compelling candidate for the SC/QH platform, since the zero-band gap of graphene ensures Ohmic contacts for most metals, including SCs with high upper critical fields. Highly transparent SC/graphene interfaces have been demonstrated with strong superconducting proximity interactions and Josephson couplings[15,17-23]. In addition, high mobility hBN-encapsulated graphene channels exhibit integer and fractional QH effects[24,25] at much lower magnetic field than the upper critical field of a few select SCs.

The microscopic picture of charge flow across the SC/QH system can be described by extending the Andreev process[1-3] between a SC and a normal conductor. Figure 1a depicts local Andreev reflection at zero magnetic field. An electron with energy ($eV$), which is smaller than the superconducting gap ($\Delta$), can enter the grounded SC electrode to form a Cooper pair, while retro-reflecting a hole back to the source of the electron. In the presence of a magnetic field, however, the chiral nature of the QH edge state forces the converted hole to keep flowing with the same chirality as the incoming electron. Note that the hole has the same chirality as the electron, because the sign of its charge and mass are both opposite to those of the electron. There can be two different regimes in this Andreev process, depending on the width of the SC electrode ($W$) compared to the superconducting coherence length ($\xi_s$). When $W \gg \xi_s$, the electron and hole will propagate along the edge of the SC electrode forming an Andreev edge state (AES)[1-3] (Fig. 1b). In a quasi-classical picture, the AES is an alternating skipping orbit of electrons and holes. Unless the probability of each Andreev reflection ($P_{AR}$) is very close to zero or unity, the AES quickly becomes an equal mixture of electrons and holes after a few bounces (Supplementary Fig. 1). The resulting mixed AES along the ground SC (i.e. at zero chemical potential) carries the averaged chemical potential equal to zero. However, when $W \ll \xi_s$ the converted hole can tunnel through the SC electrode and continue to flow on the other side of the QH edge state, a process termed as crossed Andreev conversion (CAC)[7] (Fig. 1c). Note that in the CAC process, the converted hole in the downstream carries negative (electron) chemical potential. Therefore, unlike the case of the AES, the spatial separation of electrons and holes in CAC facilitates their independent detection by measuring the chemical potential of upstream and downstream edge states with respect to the ground SC electrode, which we demonstrate in this paper. Interestingly, in the asymptotic limit of $W \ll \xi_s$ and a long finger-like SC electrode whose length $L \gg h v_F/\Delta$, the CAC picture corresponds to creating two non-Abelian anyons in resonance with the QH edge state[7-12]. Here, $h$ is a Planck constant and $v_F$ is the Fermi velocity of QH edge states. The counter propagating QH edge modes along the both sides of the finger electrode are coupled by the SC, creating two zero modes: one in resonance with the QH edge and the other at the end of the SC finger (inset of Fig. 1c). In order to measure the chemical potential of QH edge states across the SC electrode, we fabricate a multi-terminal graphene device with a SC drain electrode as shown in Fig. 1d. To obtain a high-quality graphene channel, we encapsulated mechanically exfoliated graphene samples with two h-BN crystals using a dry-transfer technique[26]. NbN was chosen for the SC drain contact, since its high upper critical field ($B_{c2} \sim 25$ T) and high critical temperature ($T_c = 12$ K) enable us to experimentally access a wide range of magnetic fields where superconductivity and the QH effect in graphene coexist.

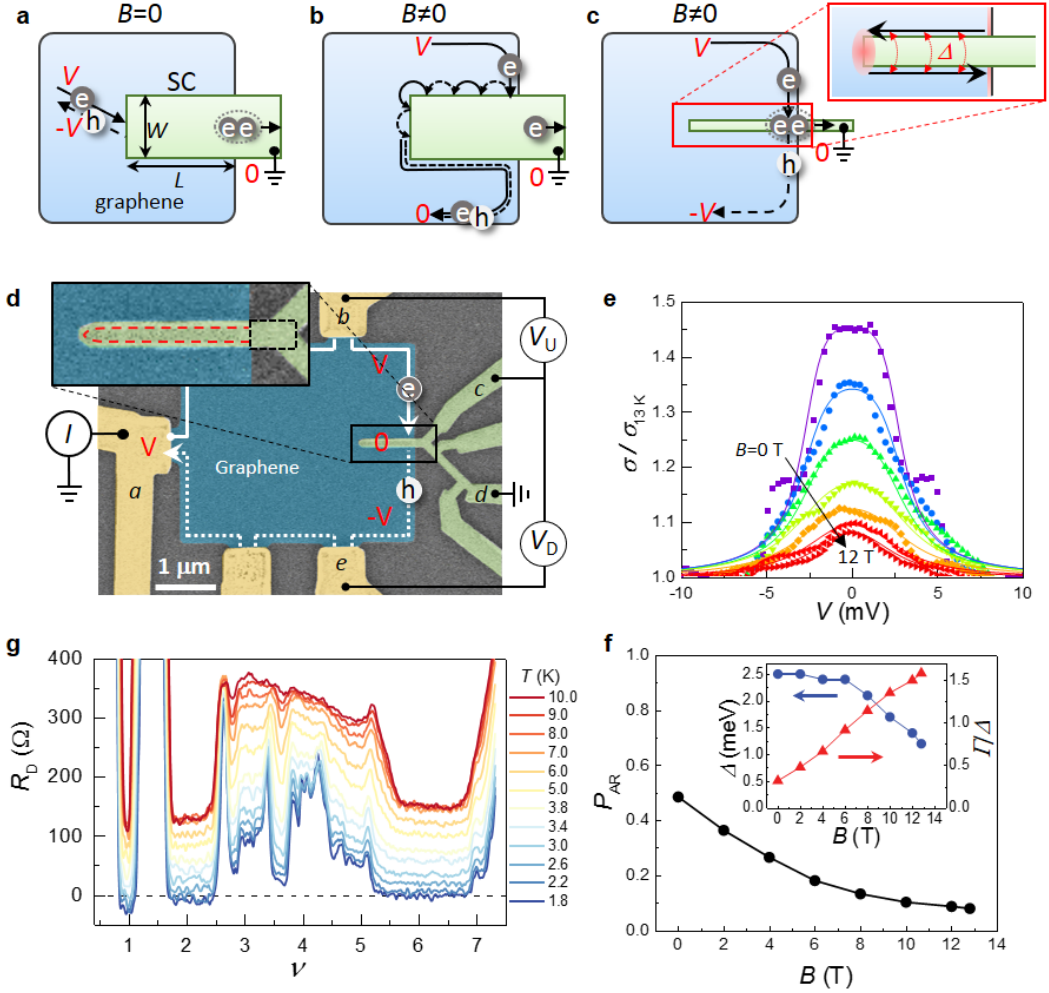

FIG. 1. **Magnetic field dependence of Andreev reflection. a-c**, Schematic diagram of an Andreev pairs of electrons and holes without an external magnetic field (**a**), and in the quantum Hall (QH) regime with a wide superconductor (**b**) or a narrow superconductor (**c**). Inset of **c**, counter-propagating QH edge states are coupled via the superconducting gap ($\Delta$), resulting in non-Abelian anyonic zero modes at the ends of the NbN electrode (red ellipse). **d**, False colour scanning electron microscopy (SEM) image of the device with measurement configurations. Ti/Au normal electrodes (yellow) and a NbN superconducting electrode (green) contacts the graphene Hall bar (blue). Inset, one-dimensional NbN contact to the graphene edge is highlighted with a dotted red line. Note that due to the finite slope of the etching profile of h-BN, the NbN contact is positioned slightly more inwards than the boundary shown in SEM image (see the schematic of cross-section in the lower inset of Fig. 4). A black dotted rectangle guides the NbN segment contributing to the downstream resistance. **e**, The differential conductance of the graphene/NbN interface at $V_{bg} = 60$ V, under various magnetic fields. The data taken at T = 1.8 K is normalised by the data taken at T = 13 K when the NbN loses superconductivity. Experimental data (symbols) are fitted to the modified BTK theory (solid lines) with a fixed barrier parameter Z = 0.18. **f**, The Andreev reflection probability ($P_{AR}$) at zero bias was calculated from the fitting parameters (inset) obtained by fittings in **e**. **g**, The filling fraction ($\nu$) dependence of the downstream edge resistance ($R_D$) at different temperatures with B = 14 T.

The contact transparency of the SC/graphene interface can be characterized by measuring the differential conductance enhancement which directly represents local Andreev process (LAP) efficiency. Fig. 1e shows the differential conductance ($\sigma = dI_{ad}/dV_{ec}$) of the graphene/NbN interface normalised by the value obtained at the temperature above $T_c$. Here $I_{ad}$ is the current biased from $a$ to $d$, and $V_{ec}$ is the voltage drop between electrodes $e$ and $c$ (marked in Fig. 1d). We apply a large backgate voltage of $V_{bg} = 60$ V to put the graphene channel in high filling fractions, away from the well-developed QH regime even at the highest field of $B = 14$ T. In this weak field limit, the magnetic field dependent $\sigma$ is determined only by the LAP, not by the modulating density of state (DOS) of the graphene (see Supplementary Fig. 2 for the case when $\sigma$ is mostly determined by the DOS of graphene in the strong field limit). Thus, $\sigma$ serves as a useful probe for characterizing the magnetic field dependence on LAP. At zero magnetic field, we observe that the normalised $\sigma$ exhibits a pronounced zero-bias enhancement up to 45%, which shows that the SC/graphene junction is reasonably transparent. As $B$ increases, the height of the zero-bias conductance peak reduces, reflecting a reduction of the Andreev reflection probability ($P_{AR}$). This reduction can be understood, considering the excitation of low-energy quasi-particles inside of the SC gap by the magnetic vortices induced by the external magnetic field. The presence of these sub-gap states allows incident electrons to enter the superconductor without the Andreev process. The Blonder-Tinkham-Klapwijk (BTK) theory that includes a magnetic field[27] fits the data with fitting parameters of $\Delta$ and energy broadening ($\Gamma$), as plotted in the inset of Fig. 1f. Here, $\hbar/\Gamma$ represents the finite quasi-particle lifetime with $\hbar = h/2\pi$. Encouragingly, $P_{AR}$ estimated from the fitting parameters is larger than 0.1 for all experimental values of $B < 14$ T.

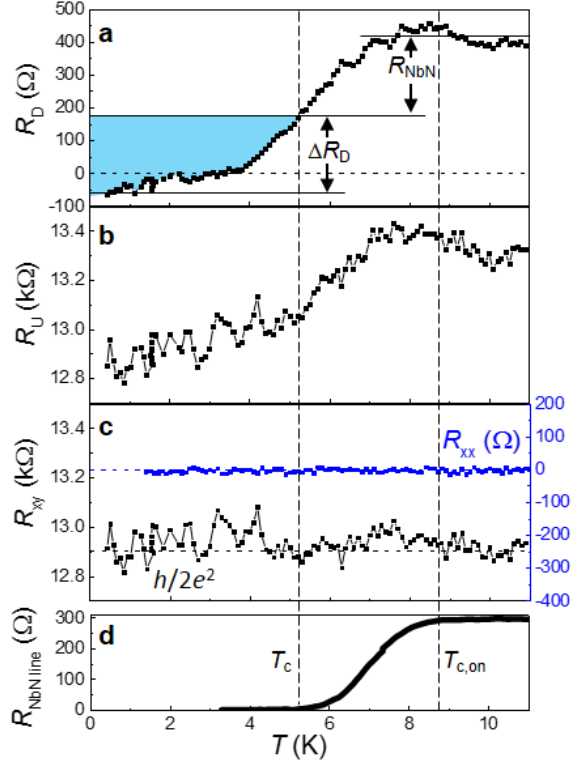

FIG. 2. **Temperature dependence of quantum Hall edge chemical potentials at $B = 8$ T. a**, The downstream resistance ($R_D$) at $\nu = 2$ decreases with temperature and eventually becomes negative. **b**, The upstream resistance ($R_U$) also changes in a similar manner to $R_D$. **c**, The quantized Hall resistance ($R_{xy} = R_U - R_D = h/2e^2$, black symbols) and vanishing longitudinal resistance ($R_{xx}$, blue symbols) confirm that the quantum Hall state is stable up to 11 K. **d**, The temperature dependence of the NbN electrode resistance ($R_{NbN\,line}$) exhibits an onset critical temperature $T_{c,on} = 8.7$ K and a critical temperature $T_c = 5.2$ K below which superconductivity in the SC electrode is fully attained.

Employing a highly transparent SC/graphene contact, the CAC process can be now investigated in dissipationless chiral edge modes in the QH regime of graphene. For this experiment we performed non-local measurements, monitoring the voltage drop between electrodes $e$ and $c$ (downstream chemical potential, $V_D$) with current $I$ biased between electrodes $a$ and $d$. There is no apparent current flow between $e$ and $c$ except along a small segment of the NbN electrode (marked with a black box in the inset of Fig. 1d), which does not contribute to the voltage drop when it is superconducting. In this non-local configuration, the measurement of $V_D$ is free from any local signal that comes from the modulation of the DOS of graphene in the QH regime, and is only sensitive to the Andreev converted holes. Figure 1g shows the downstream resistance ($R_D = V_D/I$) as a function of the filling fraction $\nu$ tuned by $V_{bg}$. The most striking feature in this graph is a negative downstream chemical potential at low $T$ for well-developed QH states of $\nu = 1, 2,$ and 6. In the Landauer-Büttiker picture this negative chemical potential corresponds to a contact transmission coefficient of the drain electrode greater than unity, that is hard to be interpreted without considering the CAC due to the SC electrode (see Supplementary Fig. 3 for the comparison with the similar device without a SC electrode). Non-local transport with negative resistance clearly indicates the presence of crossed Andreev converted holes, namely the coupling between two QH edge states via the narrow SC electrode. Negative $R_D$ is consistently reproduced in five other devices (see Supplementary Fig. 4). However, when the current flows through the bulk ($\nu \neq 1, 2$ or $,6$) $R_D$ recovers its positive value.

To further quantitatively analyse the CAC, we now discuss the temperature dependent negative downstream chemical potential for the $\nu = 2$ QH state. Its large Landau energy gap (~ 100 meV at $B = 8$ T) allows us to study the system below and above $T_c$ without degrading the QH edge state. As shown in Fig. 2a, $R_D$ for the $\nu = 2$ QH state starts to decrease as $T$ decreases only for the range below the onset critical temperature ($T_{c,on} = 8.7$ K). Note that in this temperature range where the superconductivity is not yet fully developed, $R_D$ still contains the positive contribution ($R_{NbN}$) which stems from a finite resistance of the NbN electrode segment due to the device geometry. The contribution from $R_{NbN}$ vanishes in the lower temperature range $T < T_c = 5.2$ K. We note that the positive value of $R_D(T = T_c) = 180$ Ω corresponds to the additional series resistance of the SC/graphene channel due to the imperfect contact. By comparing $R_D$ at $T > T_{c,on}$ and $T = T_c$, we determine $R_{NbN} \sim 250$

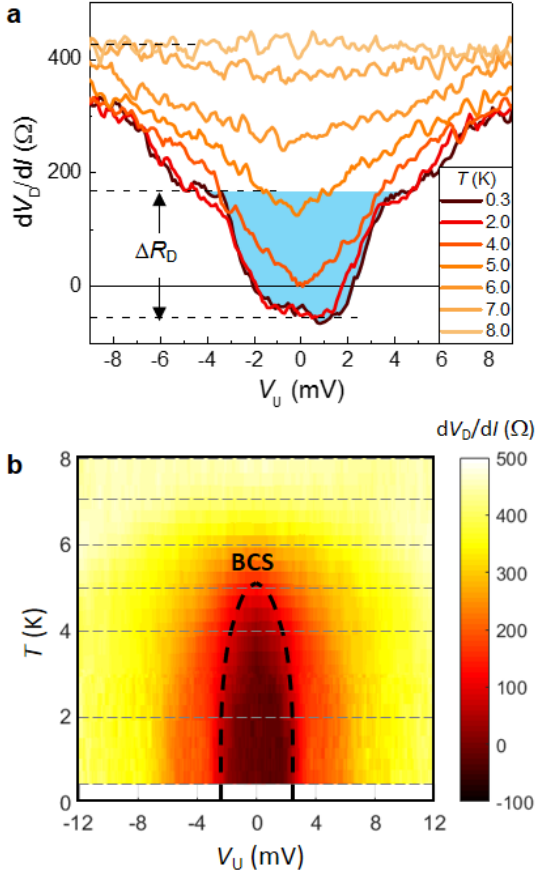

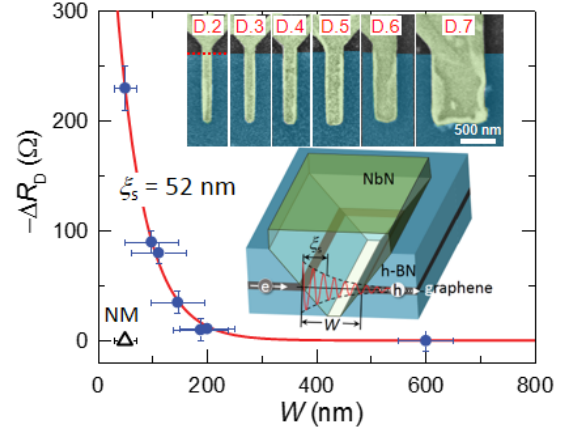

FIG. 4. **Superconducting electrode width ($W$) dependence on negative response.** The downstream resistance change ($\Delta R_D$) exponentially decreases as W increases, due to the suppression of the CAC in a wide SC electrode. The data is fitted to the exponential function of $\Delta R_D = \Delta R_{D,0}\exp[-W/\xi_s]$, with the superconducting coherence length ($\xi_s$) of NbN and the zero-width-limit value ($\Delta R_{D,0}$) as fitting parameters. Upper inset, False-coloured scanning electron microscope images of the devices of $W$ = 98, 111, 146, 188, 200, 600 nm, from the left to the right, respectively. Lower inset, A detailed schematic of the cross-section along the dotted red line in upper inset. Owing to the finite slope of the etching profile of the top h-BN, the effective width ($W$) between two graphene/NbN contacts is smaller than the apparent width of the superconducting electrode measured by the scanning electron microscope on the order of top h-BN thickness.

FIG. 3. **Bias dependence of the negative response. a**, The differential downstream resistance ($dV_D/dI$) as a function of the incident electron voltage (upstream voltage, $V_U$), at different temperature values. $dV_D/dI$ becomes positive when $|eV_U| > \Delta$, with the resistance change of $\Delta R_D = -230\,\Omega$ from the zero bias value. **b**, A colour-coded plot of $dV_D/dI$ shows the temperature dependence of $\Delta$, which follows the temperature dependence of the BCS theory (dotted line).

$\Omega$, nearly insensitive to temperature $T > T_{c,on}$. This estimation is also in good agreement with the value estimated from the NbN resistance measured from a segment fabricated as a part of the SC electrode in Fig. 2d. While the negative value of $R_D$ is clearly visible without any further analysis, our quantitative analysis of the finite positive contribution indicates that the full contribution from CAC process is considerably larger than the bare value of $R_D$. We define the net contribution of CAC process by $\Delta R_D = R_D(T = 0.3\,\text{K}) - R_D(T = T_c)$, which is represented by the shaded area in Fig. 2a. At the lowest temperature of our experiment 0.3 K, $\Delta R_D = -230\,\Omega$, about 4 % of the signal expected for perfectly efficient CAC process ($\Delta R_{D,\text{max}} = -h/4e^2$), where $e$ is an electron charge. It is also interesting to note that not only the downstream but also the upstream chemical potential ($V_U$) changes as the drain electrode becomes superconducting in our constant current-bias scheme. Fig. 2b shows the upstream resistance ($R_U = V_U/I$) as a function of temperature. $R_U(T)$ behaves similarly to $R_D(T)$ except for the constant offset independent of the temperature. The difference between $R_U$ and $R_D$ is directly related to the Hall resistance. In our current-biasing scheme, the total conductance ($\sigma = I/V_U$) enhancement by Andreev conversion processes reduces $V_U$ to keep $I$ constant. Therefore, $V_U$ decreases as much as $V_D$ does, and the Hall resistance ($R_{xy} = R_U - R_D$) stays precisely at $h/2e^2$ although $R_U$ and $R_D$ are changed by the CAC (Fig. 2c).

In principle, the observed negative non-local signal in $R_D$ can possibly be also attributed to other effects, such as ballistic electron transport[26] or viscous electron backflow[28]. To rule out these non-SC related scenarios, we further examined the differential resistance ($dV_D/dI$) as a function of the measured bias voltage $V_U$ (Fig. 3). When the incident electron energy ($eV_U$) exceeds the SC gap ($\Delta$), electrons can enter the quasi-particle state above the SC gap and the Andreev process probability is rapidly suppressed. As shown in Fig. 3, $dV_D/dI$ sharply turns from a negative value (-50 $\Omega$) to a positive value (180 $\Omega$) as $V_U$ exceeds the crossover voltage ~2 mV. This crossover voltage agrees well with $\Delta/e$, estimated from the independent measurement in the inset of Fig. 1f. We note that the amount by which $dV_D/dI$ sharply increases across the crossover voltage agrees with the value of $\Delta R_D$, independently determined in Fig. 2a. Above this crossover voltage, a broad background develops in $dV_D/dI$. As we further increase $V_U$ above 5 mV, $dV_D/dI$ approaches the value of $R_D$ at $T > T_{c,on}$ where the superconductivity of the NbN electrode completely vanishes. Indeed, the colour-coded plot of $dV_D/dI$ in Fig. 3b exhibits a BCS-type temperature dependence on $\Delta$, indicating that $R_D$ is closely related to the superconductivity of the drain electrodes.

The CAC process with different widths $W$ allows us to study its tunnelling nature and to estimate the SC coherence length $\xi_s$. During the Andreev process, a pair of electrons propagates through the superconductor as an evanescent mode, eventually condensing into a Cooper pair within the length scale of $\xi_s$ (lower inset of Fig. 4). For a SC electrode with $W \lesssim \xi_s$, the converted hole has a finite probability of tunnelling through the SC and continuing on to the other side of the QH edge state. Since the evanescent pair wave function decays exponentially, the CAC process is suppressed as $W$ increases. To quantify this exponential suppression of the CAC process, we study the $\Delta R_D$ found in devices with different $W$, in the range of 50 to 600 nm. Figure 4 shows $\Delta R_D$ versus $W$, obtained from seven different devices. All devices with $W \lesssim 300$ nm exhibit a finite negative non-local signal, which rapidly decreases as $W$ increases. We fit data to $\Delta R_D(W) = \Delta R_{D,0} \exp[-W/\xi_s]$, where $\Delta R_{D,0}$ and $\xi_s$ are the two fitting parameters corresponding to the maximal CAC efficiency in a zero-width limit and the SC coherence length, respectively. From fitting to the data, we obtain $\xi_s = 52 \pm 2$ nm, which approximately agrees with the SC coherence length of NbN, $\xi_{BCS} \sim \hbar v_F/\pi\Delta \sim 100$ nm. We note that $\Delta R_{D,0} = -600 \pm 27$ $\Omega$ is still well below the ideal value of $\Delta R_{D,max} = -h/4e^2$, presumably due to the presence of the magnetic field induced quasi-particle excitations in the SC. The low $\Delta R_{D,0}$ can be related to the degraded Andreev reflection efficiency $P_{AR}$. From Fig. 1f, we obtained $P_{AR} \sim 0.1$ at $B = 8$ T. Considering the maximum efficiency is bounded by $\Delta R_{D,max}$, we take the product of $P_{AR}$ to $\Delta R_{D,max}$, resulting in $\sim -640$ $\Omega$, which is close to $\Delta R_{D,0}$. Since the magnetic field necessary for the QH effect also degrades the CAC, a careful optimization is needed to maximize the CAC process of QH edge states within the constraint.

Although we mostly focused on the most stable $\nu = 2$ QH state in this study, a negative downstream chemical potential is also clearly observed in $\nu = 1$ QH state. The spin-polarized (or spinless) $\nu = 1$ QH state is of particular interest as a possible host of Majorana zero modes. It is intriguing to note that the spin-polarized $\nu = 1$ QH state demonstrates the CAC process with a spin singlet s-wave SC, showing a substantially large negative downstream chemical potential. We understand this CAC process to be enabled by the large spin-orbit coupling inherited from the NbN superconducting electrode[29], where spin is no longer a good quantum number. The general idea and methodology presented in this work can be readily extended to $\nu = 1$ for hosting Majorana zero modes or to fractional QH states for parafermionic zero modes, opening a new route towards a universal topological quantum computation.

**Methods** The superconducting film is optimized for high transparent contact to the graphene channel. After plasma etching of the h-BN encapsulated graphene, 5-nm-thick Ti is electron-beam evaporated at a pressure lower than $1\times10^{-8}$ Torr, followed by sputtering 5-nm-thick Nb and 50-nm-thick NbN superconducting films without breaking vacuum. For NbN film, we employ reactive sputtering in an Ar/N$_2$ environment. A thin layer of Nb film is deposited before the NbN film formation to protect the initial Ti layer from the reactive N radicals during the NbN film growth. The width of the SC is limited by the fabrication process ($W > 50$ nm). The length of the SC electrode ($L \sim 1$ μm) was chosen to ensure the aforementioned asymptotic limit to create the zero-energy resonance mode at the end of SC electrode. All the other electrodes for source and voltage probes are made of Ti/Au normal metal. We fabricate the contacts by an in-situ etching scheme[21,22,30] to improve the contact transparency. It is also important to note that there is no graphene left underneath the SC electrode due to the in-situ etching, leaving a narrow trench of graphene whose edge makes electrical contact to the SC.

**Acknowledgments** We thank S.-C. Zhang, B. Halperin and J. Alicea for fruitful discussions. The major experimental work, including sample preparation and measurement are supported by DOE (DE-SC0012260). G.-H. L acknowledges support from the Nano Material Technology Development Program through the National Research Foundation of Korea (NRF) funded by the Ministry of Science,
ICT and Future Planning (2012M3A7B4049966). P.K. acknowledges partial support from the Gordon and Betty Moore Foundation's EPiQS Initiative through Grant GBMF4543 and ARO (W911NF-14-1-0638). K.H. is supported by NSF (EFRI 2-DARE 1542807). A.Y. acknowledges support from the U.S. DOE Office of Basic Energy Sciences, Division of Materials Sciences and Engineering under award de-sc0001819. D.S.W. acknowledges the support from the National Science Foundation Graduate Research Fellowship under Grant No. DGE1144152. K.W. and T.T. acknowledge support from the Elemental Strategy Initiative conducted by the MEXT, Japan and JSPS KAKENHI Grant Numbers JP26248061, JP15K21722, and JP25106006. A portion of this work was performed at the Center for Nanoscale Systems at Harvard, supported in part by an NSF NNIN award ECS-00335765.

**Author contribution** G.-H.L. and P.K. conceived the idea and designed the project. P.K. supervised the project. G.-H.L., K.-F.H., and S.H. fabricated the devices. T.T. and K.W. provided single crystal of h-BN. G.-H.L. and D.S.W. performed the measurements. G.-H.L. and P.K. analysed the

# Supplementary Information

# Inducing Superconducting Correlation in Quantum Hall Edge States

Gil-Ho Lee[1], Ko-Fan Huang[1], Dmitri K. Efetov[2], Di S. Wei[1], Sean Hart[1], Takashi Taniguchi[3], Kenji Watanabe[3], Amir Yacoby[1], and Philip Kim[1,*]

[1]Department of Physics, Harvard University, Cambridge, Massachusetts 02138, USA
[2]Department of Electrical Engineering, M.I.T., Cambridge, MA 02138
[3]National Institute for Materials Science, Namiki 1-1, Tsukuba, Ibaraki 305-0044, Japan
*Correspondence and requests for materials should be addressed to P.K. (email: pkim@physics.harvard.edu).

**S1. Andreev reflection probability under the magnetic fields**

In a quasi-classical picture, electrons and Andreev-reflected holes in the quantum Hall (QH) regime[1-3] propagate together along the superconducting interface forming a skipping orbit. This picture is valid when the crossed Andreev process is fully suppressed (i.e. when the width of the superconductor is much larger that the superconducting coherence length $\xi_s$). We can evaluate how fast impinging quasi-particles (e.g., electrons) become an equal mixture of electrons and holes by calculating the probability of getting a sign-reversed outcome (hole) for various Andreev reflection probabilities ($P_{AR}$). Fig. S1 shows quasi-classical calculation of the probability of obtaining a quasi-particle with reversed sign as a function of the number of bounce to the SC interface at given $P_{AR}$. Unless $P_{AR}$ is very close to 0 or 1, the outgoing current becomes an equal mixture of electrons and holes within 20 bounds at the interface, which corresponds to travel of less than 200 nm along the superconducting interface at $B$ = 8 T where the magnetic length is about 10 nm.

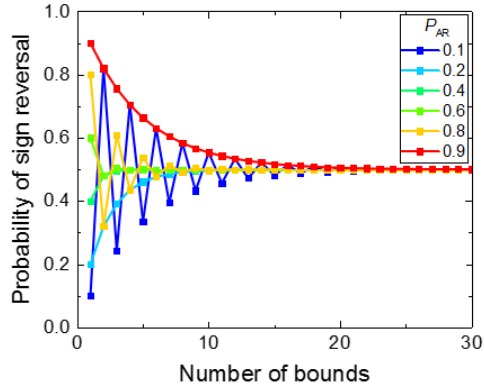

**Figure S1 | Probability of the sign reversal of incident particles.** Quasi-classical calculation of the probability of obtaining a quasi-particle with reversed sign after multiple bounds across the superconducting surface with the probability of Andreev reflection, $P_{AR}$.

**S2. Local differential conductance measurement in the magnetic fields**

Local differential conductance ($\sigma$ = d$I$/d$V$) measurements are a convenient way to investigate Andreev reflection (or superconducting proximity effect), which is well described by the Blonder-Tinkham-Klapwijk (BTK) theory[4]. Here, 'local' means that the current flows along the path of the voltage probe. In a magnetic field, however, the local differential conductance ($\sigma$ = d$I$/d$V$) is determined by the combination of Andreev reflection at the interface and the density of states (DOS) of graphene modulated by Landau level (LL) formation under the magnetic field. In the presence of LLs in the QH regime, the DOS of graphene changes dramatically with varying magnetic field ($B$) or back gate voltage ($V_{bg}$). To evaluate the contributions from the AR process and the LL DOS modulation separately, we fabricated two independent devices of bilayer graphene, each of which is contacted with superconductor (SC/G) or normal metal electrode (NM/G). Hall resistance shown in Figs. S2a and b shows well quantized plateaus for both cases. At $B$ = 0, differential conductance enhancement near zero-bias in SC/G exhibits Andreev reflection process with a transparent contact (Fig. S2c), while NM/G displays no appreciable changes (Fig. S2d). Despite of this difference at $B$ = 0, they show similar peak and dip features of $\sigma$ at higher $B$ (Figs. S2e and f) where Landau levels (LLs) are formed. This high field modulations can be understood rather in a straightforward way for NM/G; the modulating DOS of graphene in the QH regime results in such an alternating features[5]. However, the situation for SC/G device is more complicated; The Andreev reflection processes appear in $\sigma$, strongly convoluted with the modulating DOS of graphene. We find that local differential conductance measurements alone are not enough to study Andreev reflection on QH edge states in our experiment.

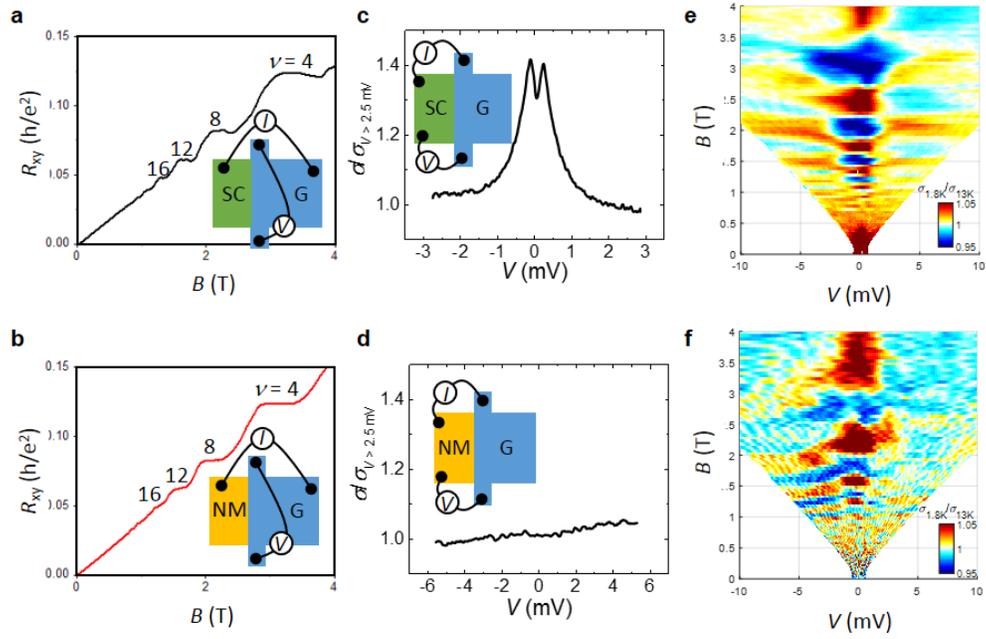

**Figure S2 | Local measurement on superconductor/graphene and normal-metal/graphene interfaces. a,b,** Magnetic field ($B$) dependence of Hall resistance ($R_{xy}$) of bilayer graphene contacted with (**a**) superconductor (SC/G) and (**b**) normal metal (NM/G) shows well quantized plateaus. **c**, Differential conductance ($\sigma$) enhancement (~ 40 %) observed with superconducting electrode shows an Andreev reflection process with a transparent contact. **d**, Normal metal contact exhibits voltage independence $\sigma$. **e,f,** Normalized differential conductance ($\sigma_{1.8K}/\sigma_{13K}$) as a function of bias voltage ($V$) and $B$ of SC/G (**e**) and NM/G (**f**). Differential conductance is normalized by the data taken at the temperature of 13 K, well above the critical temperature of superconductor.

**S3. Non-local measurement with all normal electrodes**

As a control experiment, we fabricated a device similar to the one studied in the main text only but with normal metals and performed nonlocal measurement in the same manner (Fig. S3). Downstream resistance ($R_D$) shows a positive value of ~ 29 Ω with no temperature dependence. Similarly, upstream resistance ($R_U$) stays constant so that Hall resistance ($R_{xy} = R_U - R_D$) is quantized to be $h/2e^2$. A longitudinal resistance ($R_{xx}$) also remains at ~0 Ω showing that ν=2 QH state is stable in the entire temperature range up to 11 K.

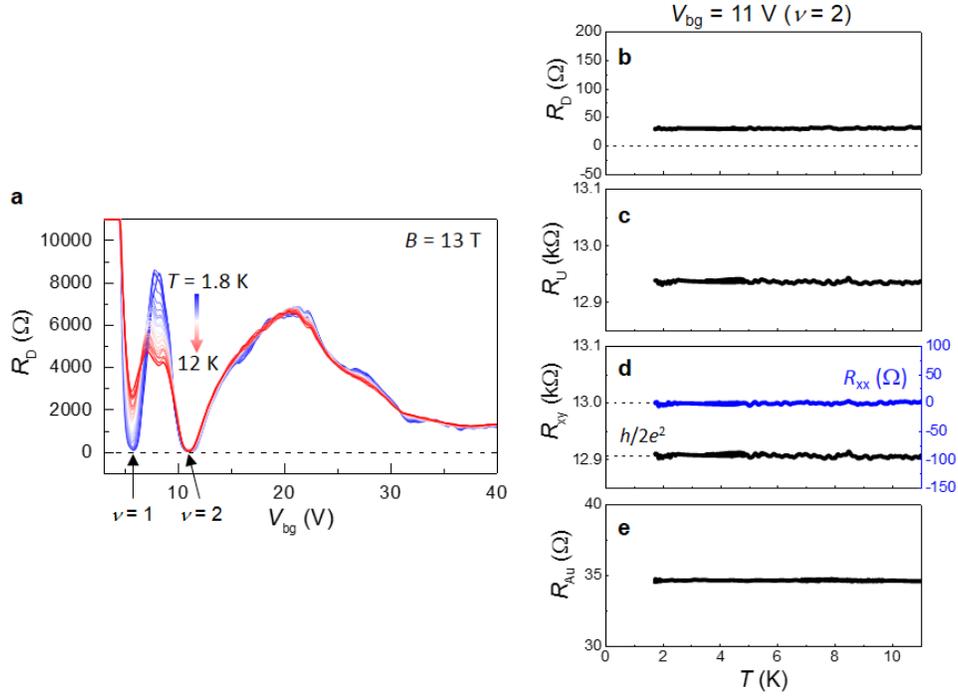

**Figure S3 | Normal drain electrode case. a**, Downstream resistance ($R_D$) in a magnetic field ($B$) of 13 T at temperature ($T$) ranging from 1.8 K to 12 K. **b-e**, Temperature dependence of (**b**) $R_D$, (**c**) upstream resistance ($R_U$), (**d**) Hall resistance ($R_{xy}$), longitudinal resistance ($R_{xx}$), and (**e**) gold electrode resistance ($R_{Au}$) shows constant behaviour within the same temperature range that was used for the experiment with superconducting contacts in the main text.

### S4. More data from other devices

The data taken from other devices of width $W = 110$ nm and $W = 600$ nm are presented in Fig. S4. As $W$ increases, the net contribution from crossed Andreev conversion ($\Delta R_D$) is considerably suppressed compared to $\Delta R_D = -230$ $\Omega$ of device with $W = 50$ nm presented in the main text.

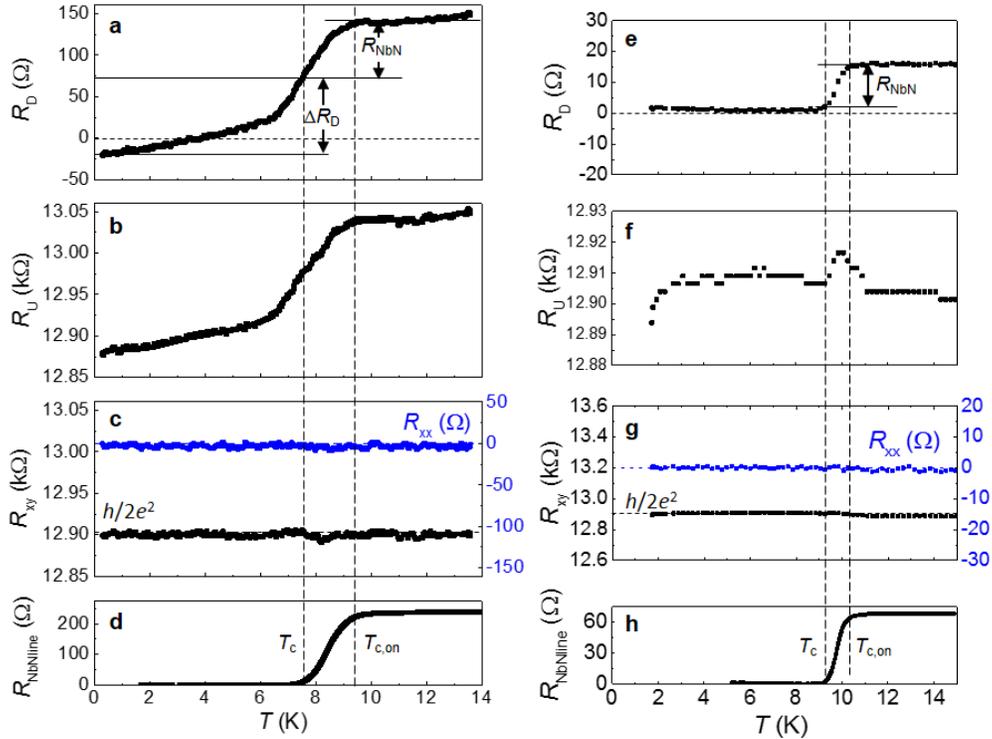

**Figure S4 | Temperature dependence of quantum Hall edge chemical potentials at $B$ = 8 T of the devices $W$ = 110 nm and $W$ = 600 nm.** Temperature dependence taken from the device of $W$ = 110 nm (**a-d**) and the device of $W$ = 600 nm (**e-h**). **a,d**, The downstream resistance ($R_D$). **b,f**, The upstream resistance ($R_U$). **c,g**, The quantized Hall resistance ($R_{xy}$ = $R_U - R_D$ = $h/2e^2$, black symbols) and vanishing longitudinal resistance ($R_{xx}$, blue symbols). **d,h**, The temperature dependence of the NbN electrode resistance ($R_{NbNline}$) shows an onset critical temperature $T_{c,on}$ and a critical temperature $T_c$ below which superconductivity of the SC electrode is fully achieved.